\documentstyle[12pt]{article}         
\begin{document}
\title{
Simulation of Beam Steering Phenomena in Bent Crystals}
\author{Valery M. Biryukov
\\  IHEP Protvino, 142284 Moscow Region, Russia
\\  E-mail:  biryukov@mx.ihep.su
\\ Presented at COSIRES 1998 (Okayama)}
\date{Published in \bf Nucl.Instrum.Meth.B153:461-466,1999}

\maketitle

\abstract{
The simulation methods for the channeling phenomena in GeV/TeV
energy range in ideal or distorted crystal lattices are
discussed. Monte Carlo predictions for feed-out and feed-in
rates, dislocation dechanneling, and deflection efficiencies
of bent crystals are compared to the experimental data.
The role of multiple interactions with crystal in circular
accelerators ("multipass channeling") for the efficiency boost
in the crystal-aided extraction experiments
is analysed.
Possible future applications of the
crystal channeling technique are considered.
}

\section{Introduction}

The experiments on charge particle beam steering by means of channeling
in bent crystals have greatly progressed
in recent years,
spanning over two decades in energy\cite{sps,e853,plb98}.
The detailed theory of beam steering is essentially based on
Monte Carlo simulations, as
tracking of a particle through a bent
crystal lattice requires not only a calculation of a particle
dynamics in this nonlinear field, but also a generation of random events
of scattering on the crystal electrons and nuclei.

To track particles
through the curved crystal lattices in simulation we apply
the approach with a continuous potential
 introduced by Lindhard.
The continuum-model description is disturbed by scattering
which could cause
the trapped particle to come  to a free state (feed out),
and  an initially free particle to be trapped (feed in).
By every step the probabilities of scattering events
on electrons and nuclei are computed depending on their local densities
which are functions of coordinates.
This ensures correct orientational dependence of all the processes
in crystal material.

\section{Numerical methods}

As the channeled particles move far from the nuclei,
the transitions between the random and channeled
states are mainly due to electronic scattering.
It is these processes which have been studied experimentally in detail.
Hence, an adequate theory needs an accurate consideration of
electronic scattering\cite{pre}.

The energy transfer $T$ in a collision
has a distribution function:
\begin{equation}	\label{dn}
\frac{d^2N}{dTdz} = \frac{D \rho_e(x)}{2 \beta^2} \frac{1}{T^{2}} \, .
\end{equation}
Here $D$=0.3071 MeV cm$^2$g$^{-1}$$Z_i^2 Z/A$; $\beta$=$v/c$;
the local density $\rho_{e}(x)$ of electrons
is normalized on the amorphous one.
Deviations  from (\ref{dn})
at $T$$\approx$$I$ or $T$$\approx$$T_{max}$ are of no concern,
due to the nature of dechanneling as explained below.
The (round) angular kick from collision is
\begin{equation}
 \theta_s = \biggl(2m_eT(1-T/T_{max})\biggr)^{1/2}/p
\end{equation}
$T_{max}\simeq 2m_ec^2 \beta^2 \gamma^2$
is the maximal transfer to a single electron.
The frequent small kicks produce a diffusion-like
angular scattering, with mean square
\begin{equation}	\label{rms}
 \theta^2_{rms} = \frac{2m_e}{p^2}
              \langle \sum_i T_i(1-T_i/T_{max}) \rangle
\end{equation}

\subsection*{Relativistic Diffusion Factor}

The diffusion approximation for {\em electronic}
scattering was used by all the authors (e.g. \cite{nitta,tar90}).
One assumes that scattering on
electrons is diffusion-like, i.e. $\theta_s \ll \theta_c$
in any collision. Then the angle of particle is changed by
small steps, with the rms value from Eq.(\ref{rms}). Note
that  $\theta_{rms}$ depends on the total
transferred energy, not on the detail of (\ref{dn}).
In the MeV range
this approximation for heavy ions works well, because even the
maximal possible angular kick per collision
$\theta_s^{max}=\sqrt{m_eT_{max}}/p\approx 1.4m_e/M$
($M$ for the particle mass) is  $<\theta_c$.

At $\sim$100 GeV (the range of modern applications
of bent crystals),  $\theta_c$ is greatly reduced to
$\sim$10 $\mu$rad, so rare catastrophic collisions with
$\theta_s$$>$$\theta_c$ may happen.
The problem for the diffusion approach is that the integration
up to $T_{max}$ in the diffusion factor
(\ref{rms}) is no longer justified.
The {\em energy} transferred with catastrophic collisions
($\theta_s$$>$$\theta_c$) is of no importance for dechanneling, and therefore
it should not be included into Eq. (\ref{rms}).
Although $\theta_{rms}^2$ depends on $T_{max}$ via $ln (T_{max}/I)$,
the removal of the energy transfers of catastrophic
collisions from  Eq.(\ref{rms}) reduces the diffusion coefficient
by a factor of $\sim$2. Hence, the diffusion approach may overestimate
the rate of feed-out and feed-in by about the same factor.
Here is an example.
For a 100 GeV proton  $T_{max}$=10 GeV.
However the transfer $T_c$ causing the angular kick (projection)
equal to $\theta_c$, is quite moderate:
\begin{equation}	\label{tc}
 T_c \approx \frac{p^2\theta_c^2}{m_e} =
          \frac{2M\gamma}{m_e} E_c
\end{equation}
$T_c$= 4 MeV for a 100 GeV proton in Si. From the energy
transferred via single collisions, about one half is due to
the scatterings with $\theta_s$$>$$\theta_c$.
For collisions with $T$$>$$T_c$ the $T$
value is no longer important. E.g., scatterings
with $T$=10 MeV and $T$=1 GeV
are equally essential as they knock the particle out of
channeling mode at once.

The above algorithm, i.e. consideration of
   electronic scattering as a series of single scattering events
   instead of the traditional diffusion approach,
 is the pronounced feature of the computer code CATCH\cite{pre}.
 Other approaches, such as analytical calculations
  by G\"{a}rtner K. et al.\cite{gaert},
 can nicely describe
   the effects dominated by the potential of crystalline lattice,
   notably the lattice distorted by defects, but the matter of our
   study -- the transitions from channeled to random states and backward
   in perfect bent crystals as induced by electronic scattering, --
   can be approached most correctly by Monte Carlo methods.

The predictions of CATCH were testified by
the recent experiments, where feeding-out, feeding-in,
high-efficiency bending, energy loss spectra, dislocation dechanneling,
and crystal extraction from accelerators have been studied \cite{pre}.

\section{Crystal extraction: multi-pass channeling}

In the extraction mode,
circulating particles may pass through the crystal many times.
Another major point: the first incidence of particles may occur
very close to the crystal edge.
The scattering of unchanneled beam in crystal, and the
accelerator optics become important.

\subsection{The SPS Experiments}

Before the CERN SPS studies of crystal extraction \cite{sps},
theoretical comparisons \cite{bi91}
with extraction experiments \cite{dubna,ass1}
were restricted by  analytical estimates only,
which gave the right order of magnitude.
The computer simulations considered idealized models only
and predicted the extraction efficiencies
always in the order of 90--99\%
while real experiments handled much smaller efficiencies,
in the order of 0.01\%.

The considered-below theoretical work has been the first
comparison between the
realistic calculation from the "first principles" (computer simulation)
and the experiment.
The simulation was performed \cite{bi78}
with parameters matching those of the SPS experiment.
Over 10$^5$ protons have been tracked
both in the crystal and in the accelerator for many subsequent
passes and turns until they were lost either at the
aperture or in interaction with crystal nuclei.
In the simulation two options were considered. The
{\em first} one assumed near-surface irregularities
(a `septum width') of a few $\mu$m.
With impact parameter below 1\,$\mu$m,
it excluded the possibility
of channeling in the first pass through the crystal.
The {\em second} option assumed perfect crystal surface.
 Table~\ref{tbs0} shows the expected extraction efficiencies
 for both options from the first simulation run
 and the measured lower limit of extraction efficiency
 as presented at the 19-th meeting on "SPS Crystal Extraction"
 \cite{meet19} held at CERN.

 \begin{table}
\begin{center}
 \caption{SPS crystal extraction efficiencies
 from the early runs, Monte Carlo and experiment} \label{tbs0}
 \begin{tabular}{lcc}
 & & \\
 \hline
 & & \\
 Option & Monte Carlo & Experiment   \\
 & & \\
 \hline
 & & \\
 Poor surface & 15\% & lower limit \\
	   &       & of 2-3\%        \\
 Ideal surface & 40\% & only known \\
 & & \\
 \hline
 \end{tabular}
\end{center}
 \end{table}

Though the efficiency comparison, theory to  measurements,
was not possible at that time,
from the analysis
one could see that the perfect-surface simulation predicted
narrow peaks for the angular scans (30 $\mu$rad fwhm) and
extracted-beam profiles, which were not observed.
The imperfect-surface option was consistent
with the experiment: wide ($\sim$200 $\mu$rad fwhm)
angular scan and sophisticated profiles of the extracted beam
(dependent on the crystal alignment).
The efficiency was then measured in the SPS
with that first tested crystal to be
10$\pm$1.7\%. Detailed simulations have shown that
efficiency should be a function of beam vertical coordinate
at the crystal,
and be from 12 to 18\% at peak, with imperfect-surface option.

\begin{figure}[htb]
\begin{center}
\setlength{\unitlength}{.9mm}
\begin{picture}(70,100)(-6,-24)
\thicklines

\put(0,-19) {\line(1,0){65}}
\put(0,-19) {\line(0,1){90}}
\put(0,71) {\line(1,0){65}}
\put(65,-19) {\line(0,1){90}}
\multiput(0,-19)(0,3.994){23}{\line(1,0){1.2}}
\multiput(0,-19)(0,19.97){5}{\line(1,0){1.6}}
\put(-7,0.99){\makebox(2,1)[l]{5}}
\put(-7,20.6){\makebox(2,1)[l]{10}}
\put(-7,40.83){\makebox(2,1)[l]{15}}
\put(-7,60.8){\makebox(2,1)[l]{20}}

\multiput(6.36,-19)(5.128,0){10}{\line(0,1){1}}
\multiput(6.36,-19)(25.64,0){3}{\line(0,1){1.5}}
\put(30,-16){\makebox(2,1)[b]{-230}}
\put(4.36,-16){\makebox(2,1)[b]{-330}}
\put(55.64,-16){\makebox(2,1)[b]{-130}}

\put(-5,73){I (rel.)}
\put(12,-24){ crystal angle ($\mu$rad)}

\put( 3.69, 4.8){ $\otimes$}
\put(16.51,13.3){ $\otimes$}
\put(24.20,21.7){ $\otimes$}
\put(29.33,57.7){ $\otimes$}
\put(34.46,19.7){ $\otimes$}
\put(42.15,14.2){ $\otimes$}
\put(54.97,12.3){ $\otimes$}

\put(23,12){70 $\mu$rad}
\put(30,19){\vector(-1,0){9.2}}
\put(30,19){\vector(1,0){8.75}}

\put(11,8){ $\star$}
\put(12,8.7){ $\star$}
\put(13,10.5){ $\star$}
\put(14,10.5){ $\star$}
\put(15,11.5){ $\star$}
\put(16,16.4){ $\star$}
\put(17,18){ $\star$}
\put(18,15.4){ $\star$}
\put(19,19.5){ $\star$}
\put(20,25.5){ $\star$}
\put(21,23.7){ $\star$}
\put(22,26.3){ $\star$}
\put(23,31.5){ $\star$}
\put(24,36){ $\star$}
\put(25,38){ $\star$}
\put(26,36.3){ $\star$}
\put(27,42){ $\star$}
\put(28,43.5){ $\star$}
\put(29,47){ $\star$}
\put(30,57.7){ $\star$}
\put(31,53.3){ $\star$}
\put(32,52){ $\star$}
\put(33,42.1){ $\star$}
\put(34,33){ $\star$}
\put(35,29.7){ $\star$}
\put(36,26.8){ $\star$}
\put(37,19.9){ $\star$}
\put(38,18.5){ $\star$}
\put(39,12){ $\star$}
\put(40,8.5){ $\star$}
\put(41,10){ $\star$}
\put(42,9){ $\star$}
\put(43,8.5){ $\star$}
\put(44,10){ $\star$}
\put(45,5.5){ $\star$}
\put(46,7.4){ $\star$}

\end{picture}
\end{center}
 \caption
  {
The angular scan of SPS extraction with a U-shaped crystal.
  Prediction ($\otimes$) and measurement ($\star$).
}
\label{u}
\end{figure}
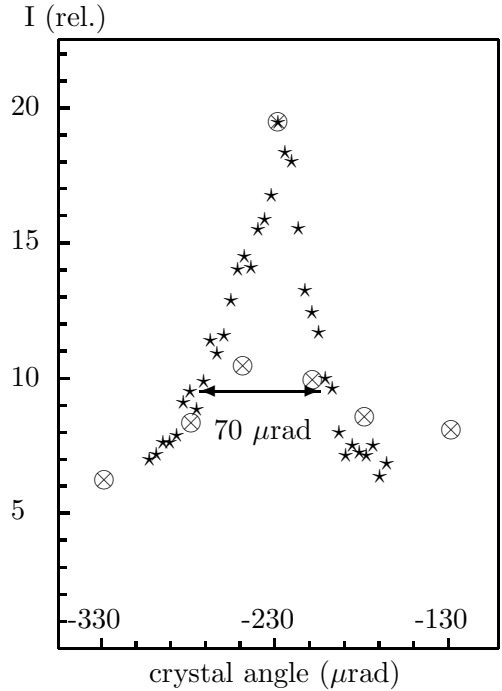

The simulation studies for a new crystal with
``U-shaped'' geometry,
performed prior to the measurements,
predicted just slight increase in efficiency
and much narrower angular scan
in the option of edge imperfection.
Fig.~\ref{u} shows one of measured scans in rather good agreement
with prediction.
Further simulation took more realistic details into model \cite{book},
and studied the dependence of crystal efficiency on septum width,
Table \ref{tab4}.
These simulations have been repeated with the energies of
14 and 270 GeV, where new measurements have been done at the SPS.
The results are shown in Table \ref{tbsps} vs measured data.

\begin{table}[htb]
\caption{
Peak efficiency $F$ vs septum width $t$.
The statistical error is 0.6\%.}
\label {tab4}
\begin{center}
\begin{tabular}{cccccc}
\hline
 & & & & & \\
 $t$ ($\mu$m) &  1  &  20  &  50  &  100  &  200 \\
 & & & & & \\
\hline
 & & & & & \\
 $F$ (\%)     & 13.9 & 12.4 & 12.9 & 10.9 &  8.2  \\
 & & & & & \\
\hline
\end{tabular}
\end{center}
\end{table}

\subsection{The Tevatron Experiment}

The Tevatron extraction experiment has provided
another check of theory at a substantially higher energy
of 900 GeV.
A detailed report of predictions for this experiment from the Monte Carlo
simulations was published\cite{pre}
two years before the measurements were taken\cite{e853}.

\begin{figure}[bth]
\begin{center}
\setlength{\unitlength}{0.75mm}\thicklines
\begin{picture}(100,55)(-50,-4)

\put(-51,0) {\line(1,0){102}}
\put(-51,0) {\line(0,1){53}}
\put(-51,53) {\line(1,0){102}}
\put( 51,0) {\line(0,1){53}}
\multiput(-50,0)(10,0){11}{\line(0,1){1.5}}
\multiput(-51,10.6)(0,10.6){4}{\line(1,0){1.5}}
\multiput(-51,0)(0,2.12){25}{\line(1,0){.5}}
\put(-40,-6){\makebox(2,1)[b]{-40}}
\put(-20,-6){\makebox(2,1)[b]{-20}}
\put(0,-6){\makebox(.2,.1)[b]{0}}
\put(20,-6){\makebox(2,1)[b]{20}}
\put(40,-6){\makebox(2,1)[b]{40}}
\put(-61,10.6){\makebox(.2,.1)[l]{0.1}}
\put(-61,21.2){\makebox(.2,.1)[l]{0.2}}
\put(-61,31.8){\makebox(.2,.1)[l]{0.3}}
\put(-61,42.4){\makebox(.2,.1)[l]{0.4}}

\put(-56,50){$F$}
\put(3,-10){$y'\; (\mu$rad)}

\put(0,30){\line(0,1){8}}
\put(-0.5,38){\line(1,0){1}}
\put(-0.5,30){\line(1,0){1}}
\put(10,28){\line(0,1){8}}
\put(9.5,28){\line(1,0){1}}
\put(9.5,36){\line(1,0){1}}
\put(20,24){\line(0,1){8}}
\put(19.5,24){\line(1,0){1}}
\put(19.5,32){\line(1,0){1}}
\put(30,19){\line(0,1){6}}
\put(29.5,19){\line(1,0){1}}
\put(29.5,25){\line(1,0){1}}
\put(40,13){\line(0,1){4}}
\put(39.5,13){\line(1,0){1}}
\put(39.5,17){\line(1,0){1}}
\put(50,8.5){\line(0,1){3}}
\put(49.5,8.5){\line(1,0){1}}
\put(49.5,11.5){\line(1,0){1}}
\put(-10,28){\line(0,1){8}}
\put(-10.5,28){\line(1,0){1}}
\put(-10.5,36){\line(1,0){1}}
\put(-20,24){\line(0,1){8}}
\put(-20.5,24){\line(1,0){1}}
\put(-20.5,32){\line(1,0){1}}
\put(-30,19){\line(0,1){6}}
\put(-30.5,19){\line(1,0){1}}
\put(-30.5,25){\line(1,0){1}}
\put(-40,13){\line(0,1){4}}
\put(-40.5,13){\line(1,0){1}}
\put(-40.5,17){\line(1,0){1}}
\put(-50,8.5){\line(0,1){3}}
\put(-50.5,8.5){\line(1,0){1}}
\put(-50.5,11.5){\line(1,0){1}}

\put(10,41.2){\circle*{1.5}}
\put(20,28.8){\circle*{1.5}}
\put(25,22.2){\circle*{1.5}}
\put(30,16.2){\circle*{1.5}}
\put(40,8.7){\circle*{1.5}}
\put(50,4.3){\circle*{1.5}}
\put(0,47){\circle*{1.5}}
\put(-10,41.2){\circle*{1.5}}
\put(-20,28.8){\circle*{1.5}}
\put(-25,22.2){\circle*{1.5}}
\put(-30,16.2){\circle*{1.5}}
\put(-40,8.7){\circle*{1.5}}
\put(-50,4.3){\circle*{1.5}}

\put(-1,34.3){$\ast$}
\put(-6,33.6){$\ast$}
\put(-11,31.6){$\ast$}
\put(-16,28.1){$\ast$}
\put(-21,23.7){$\ast$}
\put(-26,19){$\ast$}
\put(-31,14.8){$\ast$}
\put(4,33.1){$\ast$}
\put(9,31.3){$\ast$}
\put(14,28){$\ast$}
\put(19,23.7){$\ast$}
\put(24,18.9){$\ast$}
\put(29,14.8){$\ast$}
\put(39,8.4){$\ast$}
\put(49,4){$\ast$}

\put(-1,38.4){$\star$}
\put(-6,36.4){$\star$}
\put(4,37.6){$\star$}
\put(-11,34.9){$\star$}
\put(-16,30.2){$\star$}
\put(-21,26.4){$\star$}
\put(-26,19.8){$\star$}
\put(-31,15.3){$\star$}
\put(-41,8.2){$\star$}
\put(9,35.1){$\star$}
\put(14,30.2){$\star$}
\put(19,26.4){$\star$}
\put(24,20.5){$\star$}
\put(29,15.3){$\star$}
\put(39,8.2){$\star$}

\end{picture}
\end{center}
\caption {
Angular scan of the efficiency at Tevatron.
Ideal crystal ($\bullet$);
imperfect crystal:
septum width $t$=1 $\mu$m ($\star$),
$t$=50 $\mu$m ($\ast$).
Also shown is the measured peak efficiency and angular scan.
      }	\label{r1}
\end{figure}
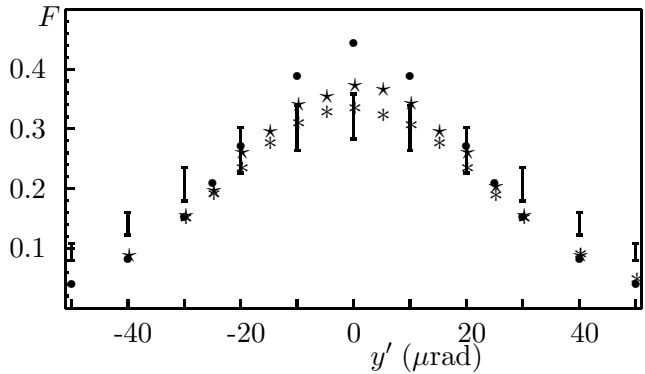

We have investigated three options:
a crystal with ideal surface, one with a septum width
(amorphous layer) of $t$=1 $\mu$m, and one with $t$=50 $\mu$m.
The  crystal bending shape and other details were as used
later in the experiment.
Figure \ref{r1} shows that there is little difference
between the three options; the peak efficiency is about 35-40\%,
and the angular scan fwhm is 50-55 $\mu$rad.
This insensitivity to the crystal surface quality is due
to the set-up different from that used in other experiments;
as a result, the starting divergence of incident protons
at the crystal was not small and hence less sensitive to edge scattering.

The measured peak efficiency was about 30\%.
This value, together with the measured angular scan,
is superimposed in Figure \ref{r1} on the theoretical expectation,
showing a rather good agreement.

\subsection{Crystal optimisation}

\begin{figure}[htb]
\begin{center}
\setlength{\unitlength}{.7mm}
\begin{picture}(80,103)(-20,-15)
\thicklines
\linethickness{.5mm}
\put(40,10){\line(0,1){5}}
\put(39,15){\line(1,0){2}}
\put(39,10){\line(1,0){2}}
\linethickness{.25mm}
\put(3,12){\circle*{2}}
\put(4,25){\circle*{2}}
\put(5,37){\circle*{2}}
\put(7,42){\circle*{2}}
\put(10,36){\circle*{2}}
\put(20,31){\circle*{2}}
\put(30,21){\circle*{2}}
\put(40,16){\circle*{2}}

\put(38.2,11.4){\small $\otimes$}
\put(8.2,28.0){\small $\otimes$}

\put(2.5,58.1){\circle*{2}}
\put(3,67.){\circle*{2}}
\put(4,72.4){\circle*{2}}
\put(5,73.6){\circle*{2}}
\put(7,75.0){\circle*{2}}
\put(10,66.3){\circle*{2}}
\put(20,56.7){\circle*{2}}
\put(30,46.5){\circle*{2}}
\put(40,37){\circle*{2}}
\put(38,33){\Large $\ast$}

\put(-10,0) {\line(1,0){60}}
\put(-10,0) {\line(0,1){85}}
\put(-10,85) {\line(1,0){60}}
\put(50,0){\line(0,1){85}}
\multiput(0,0)(10,0){6}{\line(0,1){1.5}}
\put(-.5,-6.){\makebox(1,.5)[b]{0}}
\put(9.5,-6.){\makebox(1,.5)[b]{1}}
\put(19.5,-6.){\makebox(1,.5)[b]{2}}
\put(29.5,-6.){\makebox(1,.5)[b]{3}}
\put(39.5,-6.){\makebox(1,.5)[b]{4}}
\multiput(-10,0)(0,10.6){8}{\line(1,0){2}}
\multiput(-10,0)(0,2.12){40}{\line(1,0){.75}}
\put(-19,21.2){\makebox(1,.5)[l]{0.2}}
\put(-19,42.4){\makebox(1,.5)[l]{0.4}}
\put(-19,63.6){\makebox(1,.5)[l]{0.6}}

\put(-17,72){$F$}
\put(15,-14){ $L$ (cm)}
\linethickness{.5mm}
\put(40,26){\line(0,1){12}}
\put(38,32){\line(1,0){4}}
\linethickness{.25mm}

\end{picture}
\caption {
Extraction efficiency vs crystal length
as simulated with imperfect surface.
Tevatron (top) and SPS (bottom).
Also shown is the measured efficiency at 4 cm.
      }	\label{fl2}
\end{center}
\end{figure}
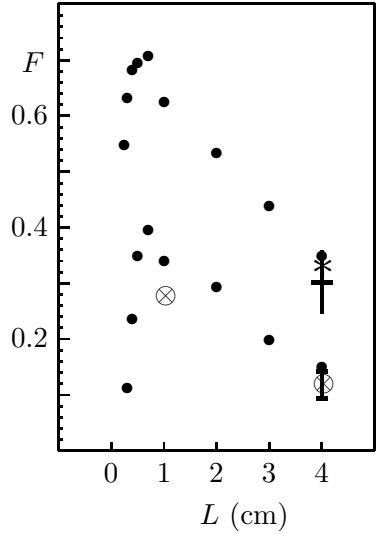

The length of the Si crystals used in the SPS and Tevatron experiments
was about optimal to bend protons with a {\em single} pass.
The efficiency  of the {\em multi}-pass extraction
is defined by the processes
of channeling, scattering, and nuclear interaction in the crystal,
which depend essentially on the crystal length $L$.
The length optimisation was a subject of a simulation, with results shown
in Fig.\ref{fl2}.

\subsection{Protvino crystal extraction experiment}

This experiment, started at the end of 1997,
aims to test crystals as short as possible
in order to gain in the extraction efficiency from
an increased number of proton encounters with  crystal,
as promised by simulations.
In the first run of 1998 this experiment established world record
of efficiency of crystal extraction, over 40\% \cite{plb98}.
The silicon crystal was just 5 mm along the beam direction,
and only 3 mm long central part was bent to give
the 70 GeV protons a deflection of 1.5 mrad.
Figs. \ref{int2},\ref{scan2} show
the extraction efficiency and
the angular scan,
Monte Carlo vs measurement.
\begin{figure}
	\begin{center}
\setlength{\unitlength}{.55mm}
\begin{picture}(110,120)(0,-3)
\thicklines
\linethickness{.3mm}
\put(     98. ,58.8)  {\circle{3}}
\put(     79. ,61.8)  {\circle{3}}
\put(     57.6 ,65.6)  {\circle{3}}
\put(     41. ,71.6)  {\circle{3}}
\put(     26.5 ,76.)  {\circle{3}}
\put(     17. ,80.6)  {\circle{3}}
\put(     9.6 ,84.)  {\circle{3}}
\put(     5.2 ,89.4)  {\circle{3}}
\put(     1.1 ,93.4)  {\circle{3}}

\put(     90. ,51.)  {\line(1,0){4}}
\put(     92. ,49.)  {\line(0,1){4}}
\put(     59.5 ,63.4)  {\line(1,0){4}}
\put(     61.5 ,61.4)  {\line(0,1){4}}
\put(     39. ,68.)  {\line(1,0){4}}
\put(     41. ,66.)  {\line(0,1){4}}
\put(     21. ,83.2)  {\line(1,0){4}}
\put(     23. ,81.2)  {\line(0,1){4}}
\put(     14. ,76.)  {\line(1,0){4}}
\put(     16. ,74.)  {\line(0,1){4}}

\put(0,0) {\line(1,0){100}}
\put(0,0) {\line(0,1){100}}
\put(0,100) {\line(1,0){100}}
\put(100,0){\line(0,1){100}}
\multiput(25,0)(25,0){3}{\line(0,1){2.5}}
\multiput(5,0)(5,0){20}{\line(0,1){1.4}}
\multiput(0,20)(0,20){4}{\line(1,0){2.5}}
\multiput(0,4)(0,4){25}{\line(1,0){1.4}}
\put(24,5){\makebox(1,1)[b]{25}}
\put(49,5){\makebox(1,1)[b]{50}}
\put(74,5){\makebox(1,1)[b]{75}}
\put(-11,20){\makebox(1,.5)[l]{10}}
\put(-11,40){\makebox(1,.5)[l]{20}}
\put(-11,60){\makebox(1,.5)[l]{30}}
\put(-11,80){\makebox(1,.5)[l]{40}}

\put(-10,105){\large F(\%)}
\put(83,-10){\large I(\%)}

\end{picture}
	\end{center}
	\caption{
Efficiency of Protvino crystal extraction as a function of the fraction
of beam store incident on the crystal;
the measurements (crosses) and simulations (open circles).
}\label{int2}
\end{figure}
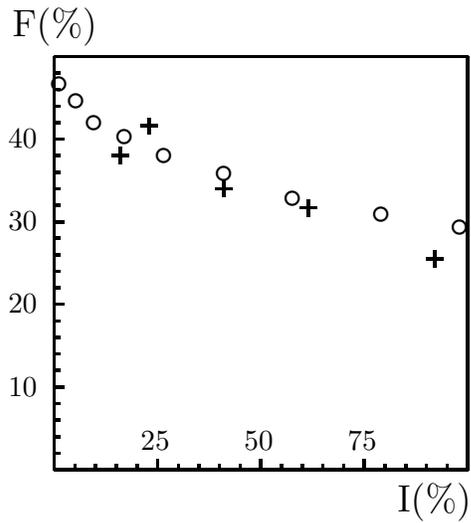
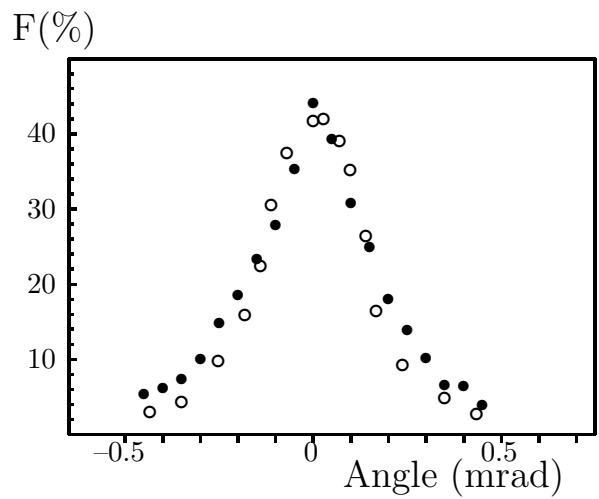
\begin{figure}
\begin{center}
\setlength{\unitlength}{.5mm}
\begin{picture}(110,150)(0,-3)
\thicklines
\linethickness{.3mm}
\put(    110.,8.)  {\circle*{3}}
\put(    105.,13.)  {\circle*{3}}
\put(    100.,13.2)  {\circle*{3}}
\put(    95.,20.4)  {\circle*{3}}
\put(    90.,27.8)  {\circle*{3}}
\put(    85.,36.)  {\circle*{3}}
\put(    80.,50.)  {\circle*{3}}
\put(    75.,61.8)  {\circle*{3}}
\put(    70.,78.8)  {\circle*{3}}
\put(    20.,10.8)  {\circle*{3}}
\put(    25.,12.4)  {\circle*{3}}
\put(    30.,14.8)  {\circle*{3}}
\put(    35.,20.2)  {\circle*{3}}
\put(    40.,29.6)  {\circle*{3}}
\put(    45.,37.2)  {\circle*{3}}
\put(    50.,46.8)  {\circle*{3}}
\put(    55.,55.8)  {\circle*{3}}
\put(    60.,70.8)  {\circle*{3}}
\put(    65.,88.2)  {\circle*{3}}

\put(    21.6,6.)  {\circle{3}}
\put(    30.,8.8)  {\circle{3}}
\put(    39.8,19.6)  {\circle{3}}
\put(    46.8,31.8)  {\circle{3}}
\put(    51.,44.8)  {\circle{3}}
\put(    53.8,61.)  {\circle{3}}
\put(    58.,75.)  {\circle{3}}
\put(    65.,83.6)  {\circle{3}}
\put(    67.8,84.)  {\circle{3}}
\put(    72.,78.)  {\circle{3}}
\put(    74.8,70.4)  {\circle{3}}
\put(    79.,53.)  {\circle{3}}
\put(    81.8,32.8)  {\circle{3}}
\put(    88.8,18.6)  {\circle{3}}
\put(    100.,9.8)  {\circle{3}}
\put(    108.4,5.6)  {\circle{3}}
\put(0,0) {\line(1,0){140}}
\put(0,0) {\line(0,1){100}}
\put(0,100) {\line(1,0){140}}
\put(140,0){\line(0,1){100}}
\multiput(15,0)(10,0){13}{\line(0,-1){2}}
\multiput(0,20)(0,20){4}{\line(1,0){2}}
\multiput(0,4)(0,4){25}{\line(1,0){1.4}}
\put(13,-7){\makebox(1,1)[b]{--0.5}}
\put(64,-7){\makebox(1,1)[b]{0}}
\put(114,-7){\makebox(1,1)[b]{0.5}}
\put(-11,20){\makebox(1,.5)[l]{10}}
\put(-11,40){\makebox(1,.5)[l]{20}}
\put(-11,60){\makebox(1,.5)[l]{30}}
\put(-11,80){\makebox(1,.5)[l]{40}}

\put(-15,105){\large F(\%)}
\put(73,-14){\large Angle (mrad)}

\end{picture}
\end{center}
\caption{
Angular scan of Protvino crystal extraction:
prediction ($\bullet$) and measurement (o).
}\label{scan2}
\end{figure}

\section{Future Applications}

To understand further possible applications of the technique
 it is useful to recall the
analytical theory of multipass crystal extraction \cite{theory}.
Few simple assumptions were taken as an input to the theory:
any particle always crosses the full crystal
length;  pass 1 is like through an amorphous matter but any
further pass is like through a crystalline matter;
that there are no aperture restrictions; particles
interact only with the crystal not a holder.

The overall multipass channeling
efficiency is then derived to be
\begin{equation} \label{fc}
F_C=\left(\frac{\pi}{2}\right)^{1/2}\frac{\theta_cx_c}{\sigma_1d_p}
   \times \Sigma(L/L_N)
\end{equation}
where
\begin{equation} \label{sum}
\Sigma(L/L_N)= \Sigma_{k=1}^{\infty} k^{-1/2}\exp(-kL/L_N)
 \simeq (\pi L_N/L)^{1/2}-1.5
\end{equation}
may be called a "multiplicity factor" as it just tells how much
the single-pass efficiency is amplified in multipasses.
A fraction (1-$G$) of the channeled particles is to be lost along the bent
crystal due to scattering processes and centripetal effects.
Then the multipass extraction efficiency is
\begin{equation}
F_E=F_C\times G=
  \left(\frac{\pi}{2}\right)^{1/2}\frac{\theta_cx_c}{\sigma_1d_p}
    \times \Sigma(L/L_N) \times G
\end{equation}
The theory check against the CERN SPS data \cite{pac97}
shows good agreement (Table \ref{tbsps}).
\begin{table}[htb]
\begin{center}
\caption{Extraction efficiencies (\%) from the SPS experiment,
 theory [17], and detailed simulations.} \label{tbsps}
\begin{tabular}{cccc}
 & & & \\
                $pv$(GeV) &    SPS     & Theory & Monte Carlo \\
\hline
 & & & \\
                14 &   0.55$\pm$0.30&  0.30 & 0.35$\pm$0.07 \\
               120 &   15.1$\pm$1.2 & 13.5 & 13.9$\pm$0.6  \\
               270 &   18.6$\pm$2.7 & 17.6 & 17.8$\pm$0.6
\end{tabular}
\end{center}
\end{table}

From (\ref{sum}) we see that multiplicity factor can be huge
if $L$ is very small or $L_N$ big.
\paragraph{MeV extraction.}
One opportunity (small $L$) is inspired by the recent successful
experiment \cite{breese} on bending 3-MeV proton beam
by means of graded composition Si$_{1-x}$Ge$_x$/Si strained layers.
This invention allows to cover the whole spectrum of accelerator
energies (from MeV to multi-TeV) by bent crystal channeling technique.
Now one can consider
extraction from accelerators starting with MeV energies,
by crystals as short as from 1 $\mu$m.
Eqs.(\ref{fc}-\ref{sum}) predict that channeling efficiency over 99\%
can be achieved in sub-GeV (and up to several GeV) range,
thus opening a new world for bent crystals applications.
With traditional bent crystals, it was common to think that highest
efficiencies
are achievable at highest (TeV) energies as multiple scattering angles
vanish with energy faster than channeling angle does.
It is very interesting now to find that channeling efficiency can be even
more boosted at lower energies due to huge multiplicity factors.
One can build a very efficient system to extract beams from
accelerators with crystals.

\paragraph{Muon extraction.}
The other opportunity (big $L_N$) for huge multiplicity factor is muons
which have formally $L_N=\infty$.
Theory then says that
efficiency of muon channeling should be 100\%.
Actually the multiplicity factor for muons is limited by
muon lifetime (mostly) and muon outscattering of the aperture.
With a muon mean lifetime of 1000 turns\cite{mox},
the number of encounters with crystal may be big.

At muon machine, the backgrounds "have the potential of killing
the concept of the muon collider"\cite{mox}.
As muons cannot be absorbed, it was proposed to extract
2-TeV muon beam halo with electrostatic septum as a primary element.
Positive muons can be easiely steered away by bent channeling crystals.
Short analysis says that we could steer negative muons also, e.g.
by the same bent planes as used for positive particles, Si(110)\cite{bau}.
In same crystal Si(110), dechanneling length $L_d$ is shortened
by factor of $\sim$100 for negative particles.
However, at 2 TeV $L_d$ is huge
($\sim$1 meter) for positives and modest ($\sim$1 cm) for negatives.
The required deflection angle is only 64 $\mu$rad[16]
and can be ensured by a Si crystal $\sim$1 mm long---quite shorter than $L_d$.
Again, multiplicity factor greatly favors muons of both sign.
One can build a very efficient system to handle halos at muon colliders.

\section*{Acknowledgements}

The author is much indebted to Professor Yamamura
and the COSIRES Organizing committee
for kind hospitality and support.

\end{document}